\documentclass[useAMS,usenatbib]{mn2e}
\usepackage{epsfig}
\usepackage{amsmath}

\newcommand{\be}{\begin{equation}}
\newcommand{\beq}{\begin{equation}}
\newcommand{\ba}{\begin{eqnarray}}
\newcommand{\ee}{\end{equation}}
\newcommand{\eeq}{\end{equation}}
\newcommand{\ea}{\end{eqnarray}}

\newcommand{\apj}{ApJ}
\newcommand{\apjl}{ApJL}
\newcommand{\mnras}{MNRAS}

\def\lsim{~\rlap{$<$}{\lower 1.0ex\hbox{$\sim$}}}

\def\gsim{~\rlap{$>$}{\lower 1.0ex\hbox{$\sim$}}}

\title[Clustering of HI galaxies]{The Halo Occupation Distribution of HI Galaxies}

\author[Wyithe et al.]{J. Stuart B. Wyithe$^1$, Michael J. I. Brown$^2$, Martin A. Zwaan$^3$, Martin Meyer$^4$\\$^1$
School of Physics, University of Melbourne, Parkville, Victoria,
Australia\\$^2$ School of Physics, Monash University, Clayton, Victoria 3800, Australia\\$^3$ European Southern Observatory, Karl-Schwarzschild-Str. 2, 85748 Garching b. M{\"u}nchen, Germany \\$^4$ School of Physics, University of Western Australia, Crawley, WA 6009, Australia\\Email: swyithe@unimelb.edu.au}

\begin{document}


\maketitle

\label{firstpage}
\begin{abstract}

  \noindent We perform an analysis of the spatial clustering
  properties of HI selected galaxies from the HI Parkes All Sky Survey (HIPASS) using the formalism
  of the halo occupation distribution (HOD). The resulting parameter
  constraints show that the fraction of satellite galaxies (i.e. galaxies which are not the central member of their host dark matter halo) among HIPASS galaxies
  is $<20\%$, and that satellite galaxies are therefore less
  common in HIPASS than in optically selected galaxy redshift
  surveys. Moreover the lack of fingers-of-god in the redshift space
  correlation function of HIPASS galaxies may indicate that the HI rich
  satellites which do exist are found in group mass rather than
  cluster mass dark matter halos. We find a minimum halo mass for
  HIPASS galaxies at the peak of the redshift distribution of
  $M\sim10^{11}$M$_\odot$, and show that less than 10\% of baryons in
  HIPASS galaxies are in the form of HI. Quantitative constraints on
  HOD models from HIPASS galaxies are limited by uncertainties
  introduced through the small survey volume.  However our results
  imply that future deeper surveys will allow the distribution of HI
  with environment to be studied in detail via clustering of HI
  galaxies.

\end{abstract}

\begin{keywords}
cosmology: large scale structure, observations -- galaxies: halos, statistics -- radio lines: galaxies 
\end{keywords}

\section{Introduction}

The cosmic star-formation rate has declined by more than an order of
magnitude in the past 8 billion years (Lilly et al. 1996, Madau et
al. 1996), a trend that is observed across all wavelengths (Hopkins 2004 and
references therein).  Why this decline has taken place, and what drives it
are two of the most important unanswered questions in our current
understanding of galaxy formation and evolution.  One of the issues
that will need to be addresses in order to answer this question is the
role of environment.  In cold dark matter cosmologies, gas cools and
collapses to form stars within gravitationally bound {\em halos} of dark
matter. These galaxies can then grow via continued star formation or
via mergers with other galaxies. As galaxies of a given baryonic mass
can only reside within dark matter halos above a particular dark
matter mass, galaxies are biased tracers of the overall dark matter
distribution. 

In linear theory, the bias in the spatial clustering of dark matter halos relative to the underlying mass distribution is a
function of halo mass but not of spatial scale. As a consequence, if
the mass power-spectrum is known, the clustering of galaxies on large
scales yields strong constraints on the masses of the dark matter
halos in which they reside. On smaller scales, the simple relationship
between galaxy clustering and halo mass breaks down. Firstly the mass
power-spectrum is in the non-linear regime. More importantly,
multiple galaxies can be distributed within individual halos (at separations $\la$1 Mpc), with the number of galaxies within halos of a given mass
exhibiting some scatter. While this complicates the modelling of galaxy
clustering, it enables measurements of spatial clustering of galaxies
to determine how galaxies populate dark matter halos as a function of
halo mass. Some understanding of these issues is provided by
simulations and these can be (and have been) tested against
observations of the spatial clustering of optically selected galaxies.

By understanding how galaxies populate dark matter halos, key insights
may be obtained into how galaxies grow over cosmic time. For example,
while the merger rate of dark matter halos is known, modelling the
dynamical friction of sub-halos (and thus galaxies) in cosmological
simulations is non-trivial and the rate of galaxy growth via merging
has been uncertain as a consequence. Knowing how galaxies populate
dark matter halos resolves this problem. In particular, consider a
case where the timescale for dynamical friction following the merger
of two dark matter halos is short compared with the Hubble time. In
this case galaxies within these halos will also merge soon after, and
satellite galaxies will be relatively rare. On the other hand, if the
dynamical friction timescale is long, then the galaxies within these
halos may remain as satellite galaxies for many Gyr. In this case
satellite galaxies will be relatively common. Brown et al. (2008)
show that the later scenario holds for the most massive dark
matter halos, with much of the stellar mass in massive halos residing
within satellite galaxies.

The way in which stellar mass populates dark matter halos has, to
first order, been determined for optically selected galaxy samples.
However little is known about how HI, the fuel for star-formation,
populates group and cluster mass dark matter halos. HI galaxies in the
Fornax region have been studied by Waugh et al.~(2002), who found
very few galaxies to be associated with the Fornax
cluster. None of the HI detections in Waugh et al.~(2002) are
early-type galaxies. Moreover, only 2 of the HI detections have both
Fornax redshifts and are within 1 degree ($\sim300$ kpc) of the
cluster centre. These results may suggest that there is a central
galaxy high mass cut-off near the Fornax cluster halo mass (which is
$7\times10^{13}$M$_\odot$ according to Drinkwater et al. 2001). 
More recently Cortese et al.~(2008) have used Arecibo to
survey a 5 square degree region around Abell 1367. They find a uniform
distribution of HI-selected galaxies throughout the volume (i.e. when
observed in HI the Abell cluster 1367 disappears), and that HI
deficiency does not vary significantly with cluster-centric
distance. These authors also find no finger-of-god effect in the 
HI-selected galaxies (in a redshift-position diagram, rather than in a
clustering analysis). Similarly, Verheijen et al.~(2007) study Abell
963 and 2192 at $z=0.2$ and find only one HI-selected galaxy within
1Mpc from the centre of each cluster. On the other hand, de Blok et
al. (2002) find that there are HI galaxies in Sculptor and Centaurus
with HI masses of $\sim10^9$M$_\odot$. However these clusters have
dynamical masses $\sim1.5$ orders of magnitude lower than that of the
Fornax and Abel clusters discussed above, and the identification of
these with the clusters is not definitive.

Thus there are many questions. For example, is
HI stripped from galaxies entering cluster, group or lower mass halos?
Is there a dark matter halo mass above which HI is heated or removed
from galaxies? Is the HI content of galaxies largely a function of
galaxy stellar mass or host dark matter halo mass? Do the stellar
masses of HI  selected galaxies grow largely via star-formation or
galaxy mergers? These questions can be addressed using the observed
clustering of HI selected galaxies to constrain models of how HI
populates dark matter halos.
A popular formalism for modeling clustering on small to large scales
is termed the halo occupation distribution (HOD) model (e.g. Peacock \& Smith~2000; Seljak
2000; Scoccimarro et al.~2001; Berlind \& Weinberg~2002;
Zheng~2004; Zehavi et al.~2004). The HOD model includes contributions to galaxy clustering
from pairs of galaxies in distinct halos which describes the
clustering in the large scale limit, and from pairs of galaxies within
a single halo which describes clustering in the small scale limit. The
latter contribution requires a parametrisation to relate the number
and spatial distribution of galaxies within a dark matter halo of a
particular mass.  Measurements of the HOD of optically selected
galaxies  provide some insights into how galaxies evolve. For
example, in the most massive dark matter halos, central galaxy stellar
mass is proportional to halo mass to the power of approximately $\sim1/3$.
Much of the stellar mass within these halos resides within
satellite galaxies (e.g., Brown et al. 2008, Moster et al. 2009). This
result implies that the mergers of dark matter halos do not always
lead to mergers of galaxies, and as a consequence massive galaxy
growth is slow relative to the rapid growth of dark matter
halos. Whether this result is also true for lower mass star-forming galaxies is
unknown at this time.

In recent years large galaxy redshift surveys such as SDSS and the 2dFGRS 
have enabled detailed studies of the clustering of in excess of
100000 optically selected galaxies in the nearby universe. 
By comparison, the largest
survey of HI selected galaxies contains $\sim5000$ sources, obtained
as part of the HI Parkes All Sky Survey (HIPASS, Barnes et al. 2001), 
a blind HI survey of the southern
sky. Meyer et al.~(2007) have studied the clustering of these HI galaxies. 
Their analysis reached the conclusion of weak
clustering of HI galaxies based on parametric estimates of correlation
length (see also Basilakos et al.~2007), but did not study the clustering in terms of the host dark
matter halo masses of the HIPASS sample. In this paper we revisit the
clustering of HIPASS galaxies using the HOD model. There are
systematic uncertainties in estimation of the observed clustering amplitude,
arising from the selection function and small survey volume (Meyer et
al.~2007), and our results show that this limits the precision with
which conclusions from clustering can be made. Nevertheless we
illustrate that the clustering of HIPASS galaxies already provides
interesting constraints on the distribution of HI within the dark
matter halo population.

\begin{figure*}
\centerline{\epsfxsize=6.9in \epsfbox{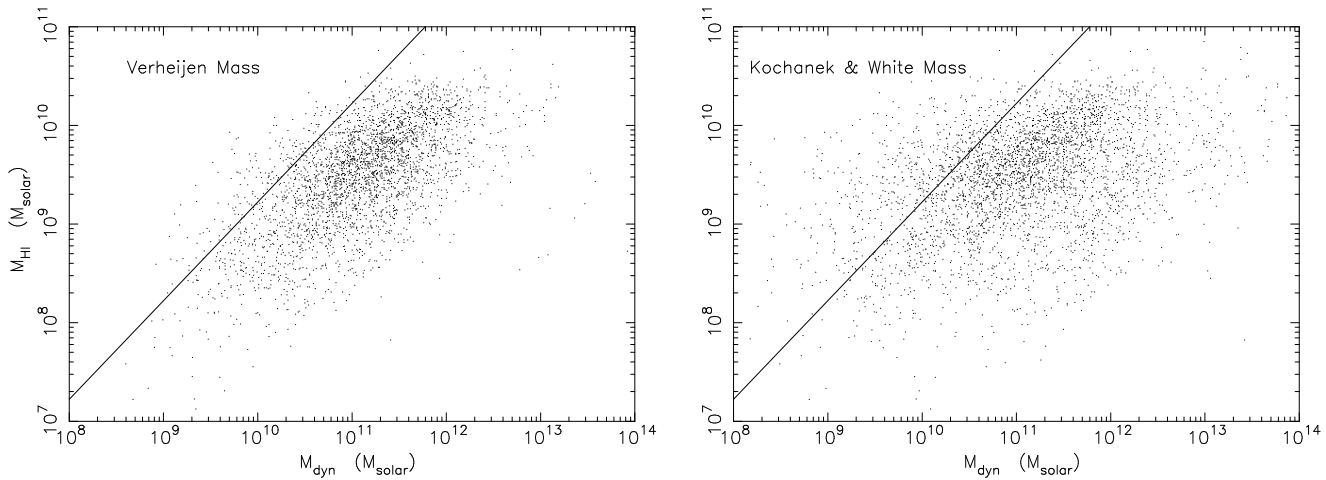}}
\caption{ Plot of HI mass verses dynamical mass for HI galaxies in HIPASS. The left hand panel plots the HI mass against a dynamical mass estimated from the circular velocity using the methods of M.~Verheijen, and by Kochanek \& White~(2001) respectively. The linear line shows the upper limit on HI mass of $M_{\rm HI}\la\Omega_{\rm b}/\Omega_{\rm M}M_{\rm dyn}$.}
\label{fig1}
\end{figure*}

This paper is organised as follows. We begin by summarising the
clustering of HI galaxies in the HIPASS survey \S~\ref{clustering}. We
then summarise the formalism for the real and redshift-space HOD
models for the correlation function (\S~\ref{HOD}), which is
discussed relative to the HIPASS observations in \S~\ref{HODobs} and
\S~\ref{HODobsz} respectively. We discuss the satelite fraction in \S~\ref{discussion} and 
summarise our findings in \S~\ref{summary}. In our numerical examples,
we adopt the standard set of cosmological parameters ~(Komatsu et
al.~2008), with values of $\Omega_{\rm m}=0.24$, $\Omega_{\rm b}=0.04$
and $\Omega_Q=0.76$ for the matter, baryon, and dark energy fractional
density respectively, $h=0.73$, for the dimensionless Hubble constant,
and $\sigma_8=0.81$ for the variance of the linear density field
within regions of radius $8h^{-1}$Mpc.

\section{clustering of HIPASS galaxies}
\label{clustering}

Meyer et al.~(2007) computed the redshift space correlation function
of HI selected galaxies from 4315 detections in the HIPASS catalogue
(HICAT; Barnes et al. 2001; Meyer et al. 2004; Zwaan et al. 2004). Correlation functions were produced by weighting each galaxy
pair equally (termed unweighted), and by weighting each pair in a way
that corrects for the survey selection function and minimises the
variance in the correlation function estimate (termed weighted). From
the redshift space correlation function, Meyer et al.~(2007) computed
the real space correlation functions in both the weighted and
unweighted cases, using inversions that were both non-parametric and
which assumed a powerlaw. 

In this paper we restrict our attention to non-parametric estimates of
the real space correlation function. However given the sensitivity of
the measured clustering to the weighting scheme adopted, we fit both
the unweighted and the weighted real-space correlation function from
Meyer et al.~(2007). From their estimated correlation function,
Meyer et al.~(2007) calculate a correlation length for the HIPASS
galaxies. In this paper our aim is instead to interpret the
astrophysical context of the measured clustering, namely the
distribution of HI galaxies within the dark-matter halo population
and the typical dark matter halo mass.

\subsection{Density of HIPASS galaxies}

Constraints on HOD models are provided both by the clustering of
galaxies, and by the density of galaxies via comparison with the
dark-matter halo mass function. We estimate the space density of
HIPASS sources from the HI mass function (Zwaan et al.~2005a), yielding
\begin{equation}
n_{\rm gal} = \theta_\star \Gamma(1+\alpha,M_{\rm HI, lim}/M_{\rm HI,\star}),
\end{equation}
where $M_{\rm HI,lim}$ is the lowest HI mass included
in the calculation of space densities, and the parameters have measured values
of $\alpha=-1.37$, $\theta_\star=0.0060$Mpc$^{-3}$, and $M_{\rm
  HI,\star}=10^{9.8}$M$_\odot$. If all HIPASS
galaxies with HI masses $>10^7$M$_\odot$ were included, the space
density would be $n_{\rm gal}\sim0.15$Mpc$^{-3}$. However HI masses of
$10^7$M$_\odot$ can only be detected out to very small distances in
HIPASS, and so are not really represented in the calculation of the
correlation function. A better estimate is obtained by looking at the
peak of the redshift distribution, where the typical HI mass is
$\sim10^{9.25}$M$_\odot$. The space density for HI masses larger than
$10^{9.25}$M$_\odot$ is $n_{\rm gal}\sim0.0069$Mpc$^{-3}$. We estimate
the error on this value to be $\sim15\%$.

\subsection{Dynamical masses of HIPASS galaxies}

An analysis of the observed clustering of a galaxy population based on
the bias of dark-matter halos implicitly assumes a relationship between
galaxy luminosity (or in this case HI mass) and the host halo
mass. Before proceeding to discuss the formalism for the model of halo
clustering we therefore describe the relation between HI mass and
dynamical mass for galaxies in the HIPASS survey. The dynamical mass
$M_{\rm dyn}$ of the HI galaxies was estimated using the circular
velocity ($V_{\rm c}$) derived from the width of the HI spectrum using
two methods. Firstly, based on the work of Marc Verheijen (PhD
thesis), we have estimated the mass of a dark matter halo with a
Hernquist~(1990) profile using the relation $M_{\rm dyn}=10^{10}R(V_{\rm
  c}/103.9\mbox{km}\,\mbox{s}^{-1})^2$M$_\odot$, where for the radius
$R$, we have adopted the B-band Kron radius (measured in kpc). 
The resulting relation is shown in the left hand panel
of Figure~\ref{fig1}. Secondly, we have also estimated $M_{\rm dyn}$ from
$V_{\rm c}$ based on Kochanek \& White~(2001), with results plotted in
the right hand panel of Figure~\ref{fig1}. Each panel includes a
linear relationship to guide the eye, showing the upper limit on HI mass $M_{\rm HI}\la\Omega_{\rm b}/\Omega_{\rm M}M_{\rm dyn}$. These panels illustrate that
while there is significant scatter, larger HI masses are found in more
massive host halos. Figure~\ref{fig1} illustrates that the
relationship between is HI and dynamical mass is shallower than
linear, with $M_{\rm HI}\propto M_{\rm dyn}^\gamma$ where
$\gamma\sim0.5-0.7$. These dynamical masses are defined such that they
are comparable to a definition based on the volume which encloses mass
at $\sim200$ times the mean density of the Universe.

The largest dynamical mass among the HIPASS sample is
$\sim10^{13}$M$_\odot$.  For comparison, we expect a number $N\sim
V_{\rm HIPASS}\times Mdn/dM=300$ of halos in the HIPASS volume $V_{\rm
  HIPASS}$, where we calculate $dn/dM$ using the Sheth-Tormen~(2002)
mass function. This yields $N\sim 300$, 30 and 1 for masses of
$M=10^{13}$M$_\odot$, $10^{14}$M$_\odot$ and $10^{15}$M$_\odot$
respectively. Figure~\ref{fig1} shows that the observed number of
these massive halos is much lower than the mass function predicts, 
although they should be detectable
throughout the HIPASS search volume. Thus it appears that the most
massive halos in the HIPASS volume do not host an HI galaxy which
traces the halo potential.

\section{HOD models}
\label{HOD}
In this paper we model the clustering of HIPASS galaxies using the
halo occupation distribution formalism (HOD; e.g. Peacock \& Smith~2000; Seljak 2000; Scoccimarro
et al.~2001; Berlind \& Weinberg~2002; Zheng~2004). Our approach is to
fit HOD parameters for the non-parametric estimate of the real space
HIPASS correlation function. Based on these fits, we then calculate
the redshift-space correlation function using the analytic formalism
described in Tinker~(2007). In this section we describe the HOD modeling formalism briefly to provide context for the particular parametrisation
used, and refer the reader to the above papers for details.

\subsection{The real-space HOD model}

The HOD model is constructed around the following simple
assumptions. First, one assumes that there is either zero or one
central galaxy that resides at the centre of each halo. Satellite
galaxies are then assumed to follow the dark matter distribution within
the halos. The mean number of satellites is typically assumed to
follow a power-law function of halo mass, while the number of
satellites within individual halos follows a Poisson (or some other)
probability distribution. 

The two-point correlation function can be
decomposed into one-halo and two-halo terms
\begin{equation}
\label{xi1}
\xi(r) = \left[1+\xi_{\rm 1h}(r)\right]+\xi_{2h}(r),
\end{equation}
corresponding to the contributions to the correlation function from
galaxy pairs which reside in the same halo and in two different halos
respectively (Zheng~2004). In real space the 1-halo term can be
computed using (Berlind \& Weinberg~2002)
\begin{eqnarray}
\label{1h}
\nonumber
1+\xi_{1h}(r)&=&\frac{1}{2\pi r^2\bar{n}_{\rm g}^2}\\
&&\hspace{-20mm}\times\int_0^\infty dM\frac{dn}{dM}\frac{\langle N(N-1)\rangle_{M}}{2}\frac{1}{2R_{\rm vir}(M)}F^\prime\left(\frac{r}{2R_{\rm vir}}\right).
\end{eqnarray}  
Here $\bar{n}_{\rm g}$ is the mean number density of galaxies. We
assume the Sheth-Tormen~(1999) mass function $dn/dM$ using parameters
from Jenkins et al.~(2001) throughout this paper. The distribution of
multiple galaxies within a single halo is described by the
function $F(x)$ which is the cumulative fraction of galaxy pairs
closer than $x\equiv r/R_{\rm vir}$. The contribution to $F$ is
divided into pairs of galaxies that do, and do not involve a central
galaxy, and is computed assuming that galaxies follow the
number-density distribution of a Navarro, Frenk \& White~(1997; NFW)
profile (see e.g. Zheng~2004). The quantity $\langle
N(N-1)\rangle_{M}$ is the average number of halo pairs. We assume an
average distribution, with $\langle N(N-1)\rangle_{M}= \langle
N\rangle_{M}^2-\langle N\rangle_{M}$.

The 2-halo term can be computed as the halo correlation function
weighted by the distribution and occupation number of galaxies within
each halo. The 2-halo term of the galaxy power-spectrum is
\begin{equation}
\label{2h}
P_{\rm gg}^{\rm 2h}(k) = P_{\rm m}(k)\left[\frac{1}{\bar{n}_{\rm g}}\int_{0}^{M_{\rm max}} dM \frac{dn}{dM} \langle N\rangle_{M} b_{\rm h}(M) y_{\rm g}(k,M)\right]^2,
\end{equation}
where $P_{\rm m}$ is the mass power-spectrum and $y_{\rm g}$ is the
normalised Fourier transform of the galaxy distribution profile
(i.e. NFW). To compute the halo bias $b(M)$ we use the Sheth, Mo and
Tormen~(2001) fitting formula. The quantity $M_{\rm max}$ is taken to
be the mass of a halo with separation $2r$. The 2-halo term for the
correlation function follows from
\begin{equation}
\xi_{2h}(r)=\frac{1}{2\pi^2}\int_0^\infty P_{\rm gg}^{\rm 2h}(k) k^2 \frac{\sin{kr}}{kr}dk.
\end{equation}

On large scales the correlation function is sensitive only to the
2-halo term, and only to the number weighted galaxy bias. However on
small scales, both the 1-halo and 2-halo terms contribute to the
clustering, and the detailed shape of the correlation function is
sensitive to the distribution of galaxies within halos. We use the
following parametrisation to describe this distribution. Halos are
assumed to host a single central galaxy and a number $N_{\rm sat}$ of
satellite galaxies if their mass is in excess of $M_{\rm min}$. The
number of satellites is taken to be a powerlaw in mass with
characteristic mass scale $M_1$ and index $\alpha$. However, motivated
by the fact that HI galaxies seem to be underrepresented as satellites
in galaxy clusters (Waugh et al.~2002), we also include an upper limit
for the halo mass which can contain an HI satellite ($M_{\rm
  1,max}$). Thus the mean occupation of a halo of mass $M$ is assumed to be
\begin{eqnarray}
\nonumber
\langle N\rangle_M &=& 1 + \langle M\rangle_{\rm sat}   \hspace{8mm}\mbox{if}\hspace{5mm}M>M_{\rm min}\\
\nonumber
&=&0\hspace{21.5mm}\mbox{otherwise},
\end{eqnarray}
where the number of satellites is defined to be
\begin{eqnarray}
\nonumber
\langle N\rangle_{\rm sat} &=& \left( \frac{M}{M_1}\right)^\alpha\hspace{10mm}\mbox{if}\hspace{5mm}M_{\rm min}<M<M_{\rm 1,max}\\
\nonumber
&=&0\hspace{21.5mm}\mbox{otherwise}.
\end{eqnarray}

\subsection{The redshift space HOD model}
\label{HODz}

Tinker~(2007) has extended the above model to calculate the redshift
space correlation function for a given HOD
parametrisation. The redshift space model is again computed based on
the sum of 1-halo and 2-halo terms as in equation~(\ref{xi1}).
In redshift space, the apparent recessional velocity of a galaxy is
the sum of its motion in the Hubble flow (directly related to its
physical distance), and of peculiar velocity (which modifies the
apparent distance based on Hubbles constant). The 1-halo term is
computed in analogy with equation~(\ref{1h}), but with an additional
integral over the line-of-sight distance and a probability
distribution for the line of sight peculiar velocity. The result of
these peculiar motions are the so-called fingers-of-god, large
line-of-sight features in redshift. Tinker~(2007)
suggests that the 2-halo term of the redshift space correlation function at apparent
line-of-sight ($r_\sigma$) and transverse ($r_\pi$) distances is most
easily computed by integrating over the 2-halo term of the real space
correlation function,
\begin{equation}
1+\xi_{\rm 2h}(r_\sigma,r_\pi)=\int_{-\infty}^\infty [1+\xi_{\rm 2h}(r)] P_{\rm 2h}(v_z|r,\phi)dv_z,
\end{equation}
where $P_{\rm 2h}(v_z|r,\phi)$ is the probability density for the
line-of-sight velocity between pairs in two distinct halos, and
$\cos{\phi}=r_\sigma/r$. Here $z^2=r^2-r_\sigma^2$, and
$v_z=H(r_\pi-z)$. Calculation of $P_{\rm 2h}$, including determination
of fitting formulae to N-body simulations is complex and we refer the
reader to Tinker~(2007) for details.

\section{Real Space HOD Models of HIPASS Galaxies}
\label{HODobs}

In this section we describe fitting of HOD models to the real-space correlation
function of HIPASS galaxies. Given the systematic uncertainty in the estimate of the correlation function owing to the small survey volume we follow Meyer et al.~(2007) and choose to fit both the unweighted and weighted HIPASS correlation
functions (although we note that the latter should more fairly estimate the correlation function that
would be obtained from a larger, volume limited survey). Our HOD model
has four free parameters, $M_{\rm min}$, $M_{1}$, $M_{\rm 1,max}$ and $\alpha$, for
combinations of which we compute the real space correlation
function, and calculate the likelihood of the model as
\begin{equation} 
\mathcal{L}(M_{\rm min},M_{1},M_{\rm 1,max},\alpha) = \exp{\left(-\chi^2/2\right)},
\end{equation}
where 
\begin{eqnarray} 
\nonumber
\chi^2&=&\sum_{i=0}^{N_{\rm obs}}\left(\frac{\log{\xi(r_i|M_{\rm min},M_{1},M_{\rm 1,max},\alpha)}-\log{\xi_{\rm obs}(r_i)}}{\sigma_{\rm obs}(r_i)}\right)^2\\
&&\hspace{5mm}+\left(\frac{\log{\bar{n}_{\rm g}(M_{\rm min},M_{1},M_{\rm 1,max},\alpha)}-\log{n_{\rm gal}}}{\sigma_{\rm gal}}\right)^2.
\end{eqnarray}
Here $\xi_{\rm obs}$ is the observed correlation function measured at
a number ($N_{\rm obs}$) of radii $r_i$, with uncertainty $\sigma_{\rm obs}(r_i)$ (in
dex), and $n_{\rm gal}$ is the observed galaxy density with
uncertainty $\sigma_{\rm gal}$ (in dex). We compute the halo density
$\bar{n}_{\rm g}$ using the Sheth-Tormen~(2002) mass function as part of the
HOD model.  The error bars on the observational estimates are not
symmetric. Note that we assume the correlation function points at
different radii to be independent (as should be the case for a small sample, with large Poisson dominated noise). Covariance between measurements of the correlation function at different radii can lead to unrealistically small errors on constrained HOD model parameters. We do not add this layer of sophistication to our analysis, owing to the large uncertainties already introduced into the clustering measurements via the chosen weighting scheme.

\begin{figure*}
\vspace{10mm}
\centerline{\epsfxsize=6.9in \epsfbox{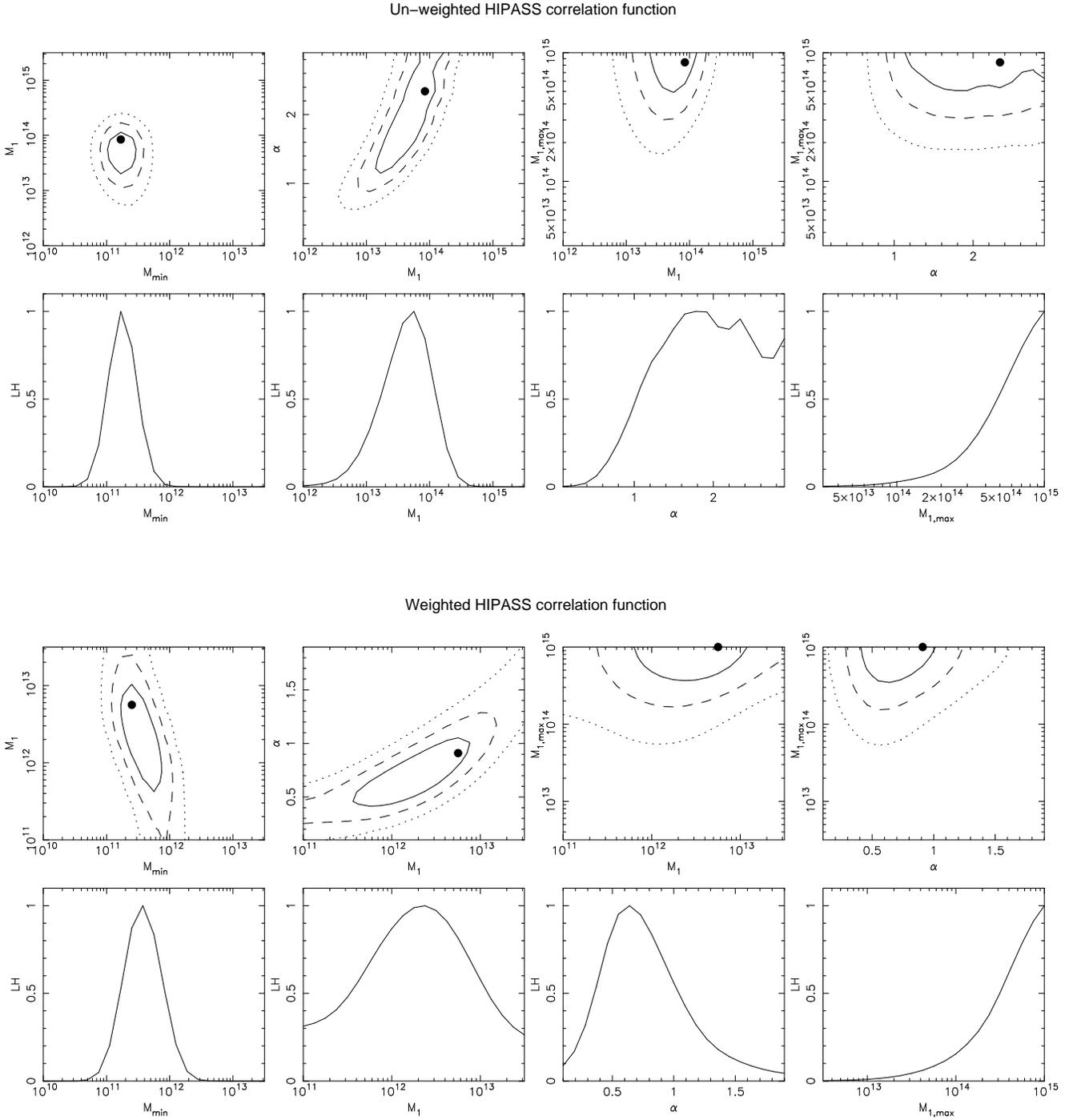}}
\caption{Constraints on HOD parameters describing estimates of the non-parametric real-space HIPASS correlation function from Meyer et al.~(2007).  Two sets of constraints are shown, based on the unweighted (upper set) and weighted (lower set) estimates of the HIPASS correlation function respectively. In each case, the \textit{Upper panels} show contours of the likelihood in 2-d projections of the 4-d parameter space used for the HOD modeling, while the {\em Lower panels} show the marginalised likelihoods on individual parameters. Here prior probabilities on $\alpha$, $\log{M_{\rm min}}$, $\log{M_{\rm 1}}$ and $\log{M_{\rm 1,max}}$ are assumed to be constant. The contours are placed at 60\%, 30\% and 10\% of the peak likelihood (the location of which is marked by a dot). }
\label{fig2}
\vspace{10mm}
\end{figure*}

We begin by fitting our HOD model to the unweighted real-space
clustering of HIPASS galaxies. The upper row of the upper set of
panels in Figure~\ref{fig2} shows contours of the likelihood in 2-d
projections of this 4-d parameter space. Here prior probabilities on
$\alpha$, $\log{M_{\rm min}}$, $\log{M_{\rm 1}}$ and $\log{M_{\rm
    1,max}}$ are assumed to be constant. The contours are placed at
60\%, 30\% and 10\% of the peak likelihood (the location of which is
marked by a dot). The lower row shows the corresponding marginalised
likelihoods on individual parameters. Meyer et al.~(2007) noted that
the correlation length of HIPASS galaxies is smaller than for optical
surveys. Here we quantify the clustering on large scales via the host
halo mass, finding a value of $M_{\rm
  min}\sim10^{11.2\pm0.2}$M$_\odot$. On smaller scales, the halo
occupation modeling illustrates the requirement of a non-zero 1-halo
term in order to reproduce the excess clustering of galaxies at $r\la1$Mpc. We find
$M_{1}\sim10^{13.6\pm0.5}$M$_\odot$, which is two orders of magnitude
larger than $M_{\rm min}$. The power-law index is
tightly correlated with $M_{\rm 1}$, but loosely constrained to be
$\alpha\ga1$. Since $M_1$ represents the characteristic mass where
satellites outnumber the central galaxies, the large value of $M_1$
indicates that there are only a small number of satellite galaxies in
the HIPASS sample. 

\begin{figure*}
\centerline{\epsfxsize=7.in \epsfbox{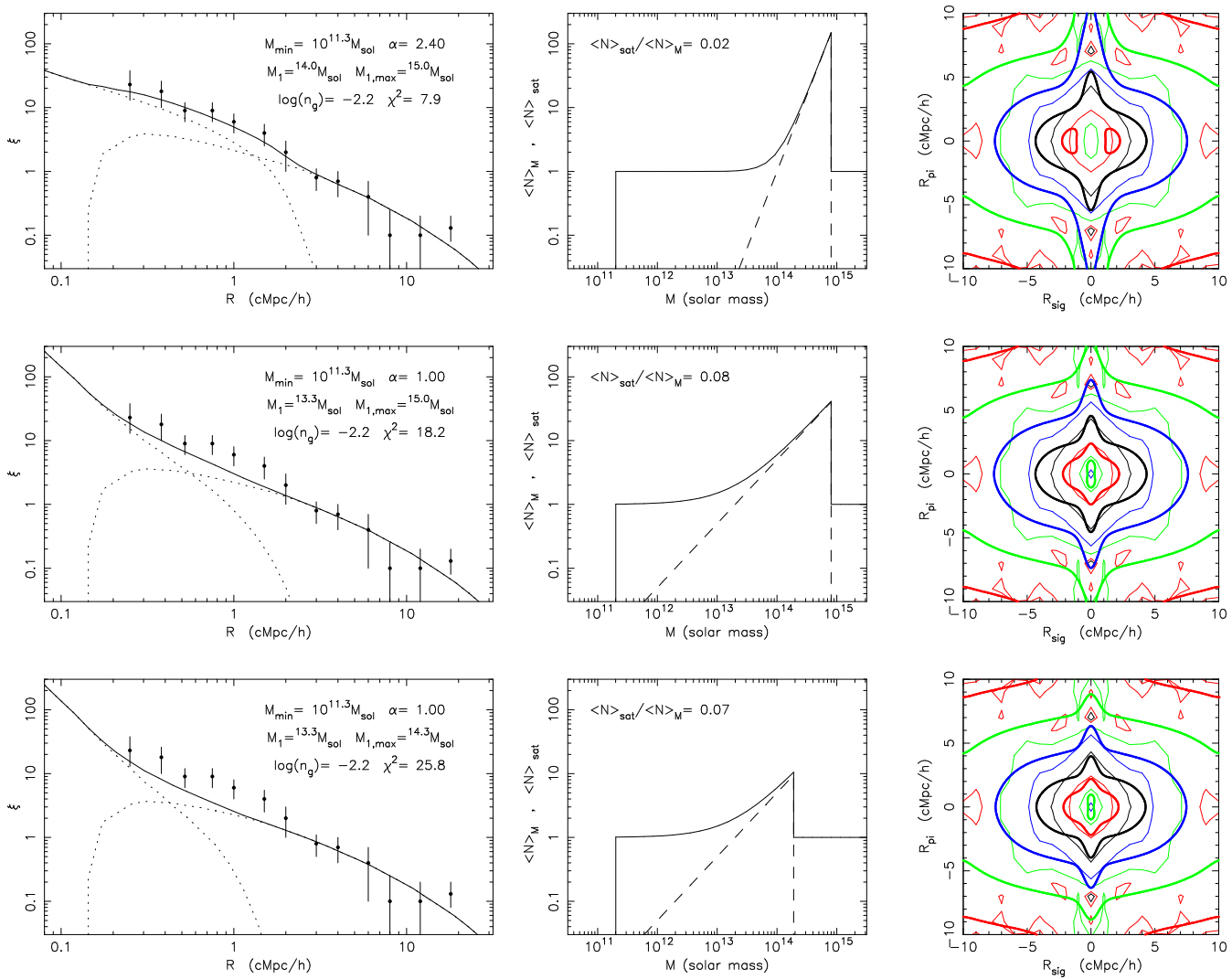}} 
\caption{
Examples of correlation functions that are consistent with the HOD parameter constraints in Figure~\ref{fig2} derived from the estimates of the unweighted real-space HIPASS correlation function. The {\em Left-hand panels} show modeled real-space correlation functions (with the 1-halo and 2-halo terms plotted as dotted curves). The non-parametric determinations of the real-space correlation function for HIPASS  galaxies are plotted as the data points with error bars. The parameters used for each model are listed together with the resulting value of $\chi^2$. The {\em Central panels} show the corresponding total (solid lines) and satelite (dashed lines) occupation numbers of galaxies as a function of halo mass.  The thick contours in the {\em Right-hand panels} show the corresponding redshift space correlation functions. The black contours correspond to $\xi=1$, with the remaining contours differing in level by factors of 2. The model correlation function has been smoothed at $0.5h^{-1}$Mpc for comparison with the data. The thin contours are the unweighted redshift-space correlation function for HIPASS galaxies (from Meyer et al.~2007).
}
\label{fig3}
\end{figure*}

\begin{figure*}
\centerline{\epsfxsize=7.in \epsfbox{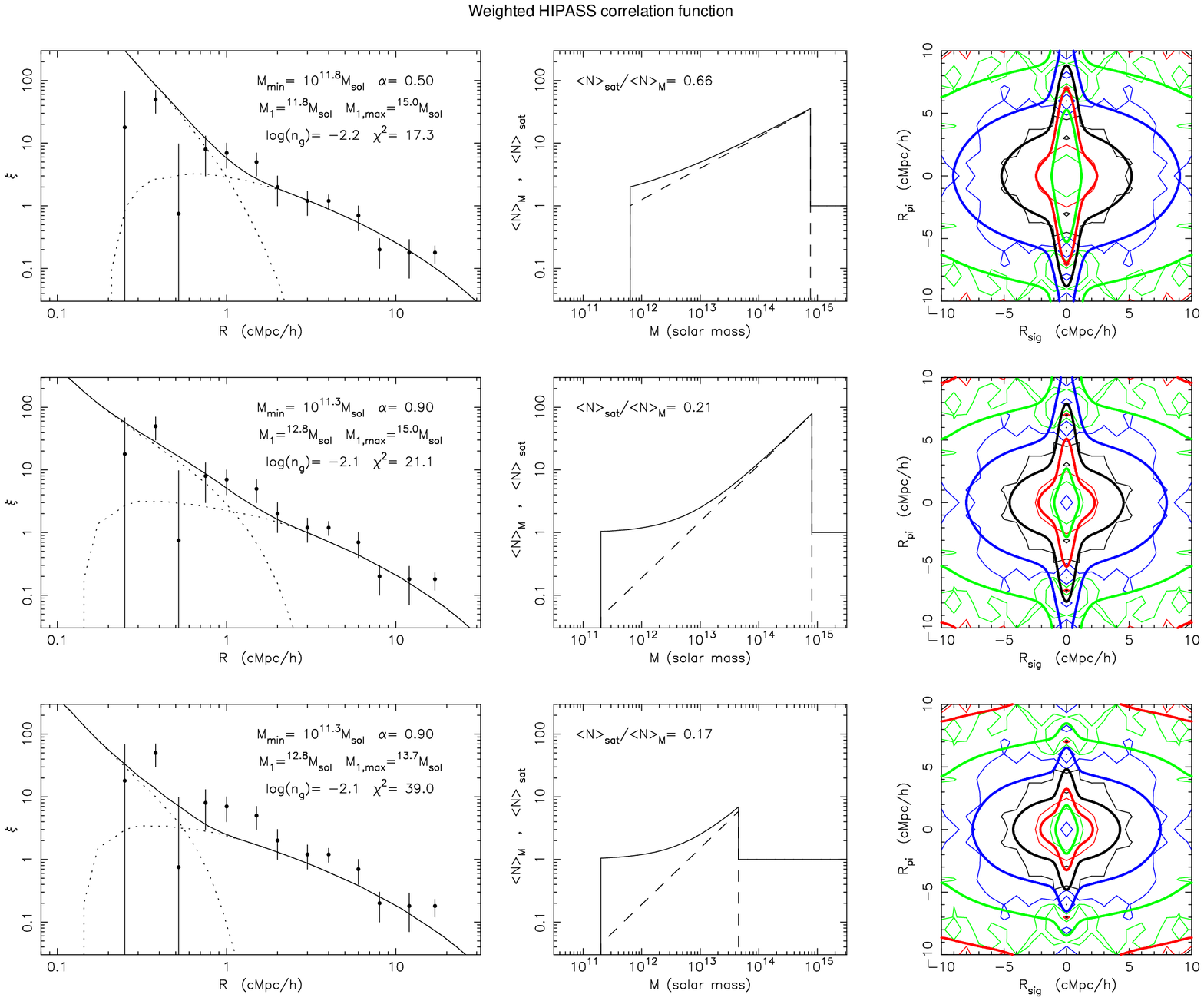}} 
\caption{
As per Figure~\ref{fig3}, but showing examples of correlation functions that are consistent with the HOD parameter constraints in Figure~\ref{fig2} derived from the estimates of the weighted real-space HIPASS correlation function.  The thin contours are the weighted redshift-space correlation function for HIPASS galaxies (from Meyer et al.~2007).
}
\label{fig4}
\end{figure*}
In the lower set of panels in Figure~\ref{fig2} we repeat this
analysis for the weighted estimate of the HIPASS real-space
correlation function. Here we find best fit estimates of
$M_{\rm min}\sim10^{11.5\pm0.3}$M$_\odot$, and $M_{\rm
  1}\sim10^{12.2\pm0.5}$M$_\odot$. There is greater tension between
the galaxy density and clustering amplitude in this case leading to
larger values of $\chi^2$ for the best fit. We find $M_{\rm
  1}\sim10M_{\rm min}$, smaller than the difference found in the
unweighted case. However the value of the power-law slope is loosely
constrained to be $\alpha\sim0.7\pm0.4$, weakly preferring satellites to
be in smaller halos (but consistent with a linear relation).  In this
case $M_1$ and $\alpha$ are again tightly correlated, with a smaller
value of $M_1$ associated with a shallower index $\alpha$ in order
to produce the low amplitude of the small scale clustering.

If $\alpha$ is forced to equal unity in our analysis, then we find
$M_{\rm 1}=10^{13}$M$_\odot\sim50-100M_{\rm min}$ for both the
unweighted and weighted estimates.  For comparison, with $\alpha=1$,
the red galaxy sample (chosen to exclude gas rich galaxies with a large star formation rate) from Brown et al.~(2008) has clustering described by 
$M_{1}\sim3$M$_{\rm min}$,
while clustering of galaxies in the Sloan Digital Sky Survey (including both gas-rich and gas-poor galaxies) suggests
$M_{\rm 1}\sim20M_{\rm min}$ (Zehavi et al.~2005). Thus the
qualitative conclusions of both the weighted and unweighted estimates
of the HIPASS correlation function are consistent; namely that HI
satellites in groups and clusters are rare compared to the results of
optical clustering studies. We return to quantify this point further
in \S~\ref{discussion}. The effect of satellites on the real space
correlation function at small scales is illustrated in Figure~5 of
Meyer et al.~(2007), where it can be seen that HI selected HIPASS
galaxies have a smaller correlation length than optically selected
samples, but also that the difference in amplitude of the correlation
function is greatest at scales less than 1Mpc, where the 1-halo term
dominates. Thus, by determining the relationship between $M_{\rm min}$
and $M_{1}$, the real space HOD correlation function quantifies
previous suggestions that HI galaxies are under-represented in
overdense environments (Waugh et al.~2002).

The inferred values of $M_{\rm min}$ for the HIPASS galaxies are
quantitatively consistent between the unweighted and weighted
clustering estimates, making estimates of the halo mass for HIPASS
galaxies fairly robust (we note that the estimates partly driven by
the galaxy density, which is common between the two cases). Moreover,
the clustering estimate of host mass from the unweighted HIPASS
correlation function is easily reconciled with the dynamical estimates
of HIPASS galaxy host masses shown in Figure~\ref{fig1}, for which the
logarithmic means are $\langle
\log_{10}(M/\mbox{M}_\odot)\rangle=11.1$ for both of the dynamical
mass estimates presented. 

\subsection{The HI mass fraction in HIPASS galaxies}
 
The halo mass estimates derived from the combination of clustering and
density of HI galaxies allow the fraction of baryonic
mass in galaxies that is in the form of HI ($f_{\rm HI}$) to be estimated. To this
end we first assume that the hydrogen to dark-matter mass ratio is the
same within galaxies as in the mean universe, so that the total
hydrogen mass within a halo of mass $M$ is $\sim \Omega_{\rm
  b}/\Omega_{\rm M}M$. We then assume that the baryon to dark matter
mass is the same for all halos, yielding
\begin{equation}
f_{\rm HI}\sim\frac{\Omega_{\rm M}}{\Omega_{\rm b}}\frac{M_{\rm HI,lim}}{M_{\rm min}}.
\end{equation}
Including the systematic uncertainty as estimated by the differing results for $M_{\rm min}$ from the unweighted and weighted clustering measurements, we find $f_{\rm HI}\sim10^{-1.4\pm0.4}$. Thus we find that less than 10\% of baryons within HI selected galaxies exists in the form of HI.

\section{Redshift Space HOD Models of HIPASS Galaxies}
\label{HODobsz}

The line-of-sight structure of the redshift-space correlation function
is dominated by gravitational infall on large transverse scales, and
by virial motions of satellites on small transverse scales. Both of
these features can be seen in the unweighted HIPASS redshift space
correlation function (plotted as the thin contours in the right hand
panels of Figures~\ref{fig3} and \ref{fig4}), though the fingers-of-god are less pronounced than expected based on optical galaxy redshift
surveys (Meyer et al.~2007). As mentioned above in
\S~\ref{clustering}, the small volume of the HIPASS survey suggests
that the correlation function should be constructed using a weighting
scheme so that it is not dominated by galaxy pairs near the peak of
the selection function. However this weighting introduces systematic
uncertainty into the determination of the correlation function. The
weighted redshift space correlation functions (plotted as the thin
contours in the right hand panel sets of Figure~\ref{fig4}) show
evidence for infall, but marginal or no evidence for fingers-of-god.

Additional information on the satellite galaxy distribution is
contained in the redshift space correlation function. In redshift
space, the line-of-sight structure of the 2-halo term is governed by
gravitational infall, while the 1-halo term is dominated by the virial
motions of satellite galaxies producing the so-called
fingers-of-god. In this section we turn to calculation of the redshift
space correlation function using the analytic HOD model of
Tinker~(2007). Given the large uncertainties in the construction of
the HIPASS correlation function, we do not fit the redshift space
correlation function directly. Rather, based on the parameter
constraints in Figure~\ref{fig2} we calculate examples of the redshift
space correlation function for qualitative comparison with the HIPASS
clustering. These examples, and their comparison with the HIPASS
redshift space correlation function, offer some hints regarding the satelite distribution that are not available from the real space correlation function alone. They also indicate the way in which the full
3-dimensional shape of the correlation function could be utilized
within a larger, more statistically representative sample.

Three examples are shown in each of Figures~\ref{fig3} and \ref{fig4}
for comparison with each of the unweighted and weighted determinations
of the HIPASS correlation function. The chosen HOD models have
parameters which adequately describe the real-space clustering, as
shown in the left hand panels. In each case the models differ in the
values chosen for various parameters. These values effect the occupation of
dark matter halos as shown in the central panels of Figures~\ref{fig3}
and \ref{fig4}. For example, smaller values of $\alpha$ and $M_1$
preferentially place the required number of satellites in smaller
halos, and so reduce the prominence of the fingers-of-god. A smaller
value of $M_1$ also lowers the typical mass at which satellites become
common, and so increases the fraction of galaxies that are satellites  (the fractions are listed in the central panels). These two
parameters are varied between the upper 2 panels of Figures~\ref{fig3}
and \ref{fig4}). In the weighted case, the models also differ in the
value of $M_{\rm min}$, with decreasing values from top to
bottom. Larger values of $M_{\rm min}$ (and hence larger values of
bias) lead to smaller values of $\beta\equiv\Omega_{\rm m}^{0.6}/b$,
and in turn to a real-space correlation function that is less
compressed along the line of sight on large transverse scales (as can
be seen in the correlation functions of Figures~\ref{fig3} and
\ref{fig4}). However the modeled fingers-of-god are more prominent
than is the case in the HIPASS data for each of these cases.

In the lower panels of Figures~\ref{fig3} and \ref{fig4} we show
examples that impose an upper limit on the host mass containing
satellite galaxies.  By excluding the presence of satellites in massive
halos, the values of $M_{\rm 1,max}=10^{14.3}$M$_\odot$ and
$M_{\rm 1,max}=10^{13.7}$M$_\odot$ in the unweighted and weighted
cases force the required number of satellites to reside in
smaller halos. This reduces the prominence of the
fingers-of-god, which are sensitive to
the magnitude of satellite virial motions within the host halo. As a result these models yield fingers-of-god which are of comparable
strength to those seen in the HIPASS data. On the other hand, these same fits to the
unweighted estimate of the real-space correlation function predict line-of-sight compressions at large transverse separations [Kaiser~(1987) effect] that appears to be too
large to explain the HIPASS data\footnote{Note that we have plotted the redshift space correlation function using reflections of the measured correlation in the first quadrant to fill the remaining three quadrants. As a result, features due to noise in the correlation function are repeated and could give the impression of a systematic difference between the shape of the data and model correlation functions where no statistically significant difference exists.}. In the weighted case the correlation function amplitude is
below the observed estimate owing to the tension between the density
and correlation function amplitudes in this case.

\section{Satelite fraction}
\label{discussion}

\begin{figure}
\centerline{\epsfxsize=2.2in \epsfbox{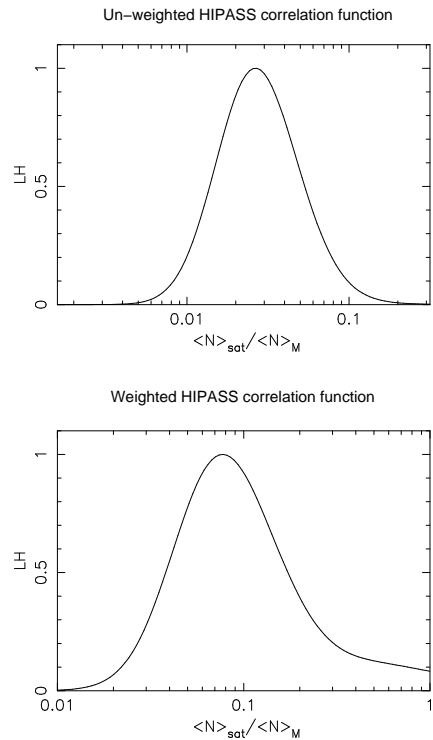}} 
\caption{
Likelihood functions (per unit logarithm) for the fraction of satellites among the HIPASS galaxies.
}
\label{fig5}
\end{figure}

Taken together the results of our modeling suggest that HI rich
satellite galaxies are not common in HIPASS, or else the 1-halo term
would be more prominent in the real-space correlation function. This
is quantified in Figure~\ref{fig5}, where we show the likelihoods (per
unit logarithm) for the ratio $\langle N\rangle_{\rm sat}/\langle
N\rangle_M$ obtained by marginalising over the HOD distributions shown
in Figures~\ref{fig3} and \ref{fig4} for the unweighted and weighted
HIPASS correlation functions respectively. A range of HOD models can
describe the HIPASS real space correlation function, and our fits
include a range of values with means near $\sim 3\%$ and $\sim10\%$
for the fraction of satellites in the unweighted and weighted cases.
Although unlikely, we find that the weighted estimate of the real
space correlation function can be described with HOD models for which
the satelite fraction is greater than 20\%. However we find that HOD
models which have more than 20\% of the galaxies as satellites have
fingers-of-god that are too prominent (e.g. see
Figure~\ref{fig4}). The satelite fraction of $\langle N\rangle_{\rm
  sat}/\langle N\rangle_M\sim0.20$ should therefore be considered an
upper limit for HIPASS galaxies.

For comparison, typical fits to the halo occupation distribution of
optical samples have satellite fractions that vary with galaxy
luminosity and type.  For example, the HOD modeling of Brown et
al.~(2008) implies a satelite galaxy fraction of $\langle
N\rangle_{\rm sat}/\langle N\rangle_M\sim 0.5$ among red galaxies with
$0.2<z<0.4$ and a comparable space density to HIPASS. This suggests
that red galaxies (which are HI poor) are more common among the
satelite population than HI selected galaxies. On the other hand, for
galaxies in the Sloan Digital Sky Survey with r-band absolute
magnitudes in excess of -19 (again a sample with a comparable density to
HIPASS galaxies) the HOD parametrization found in Zehavi et al.~(2005)
implies a satelite fraction of $\langle N\rangle_{\rm sat}/\langle
N\rangle_M\sim 0.25$. This value lies between the fraction we find
from HIPASS, and the fraction found for red galaxies (Brown et
al.~2008). Zehavi et al.~(2005) divide their galaxy population into
blue and red galaxies. They find that the red galaxy population has a
steeper correlation function, which, when interpreted in terms of the
HOD model implies that satelite galaxies are rarer among blue galaxies than among red galaxies.  Thus there appears to be a
sequence of satellite fractions. A sub-sample of red galaxies includes a larger
proportion of satellites than does a sub-sample of blue galaxies, which in turn has a larger proportion of satellites than an HI selected sub-sample of
galaxies. 

Thus, as with observations of optical galaxy clustering, studying how
HI galaxies populate dark matter halos provides important insights
into how galaxies are assembled and evolve over cosmic time. For
example, if massive galaxies grow largely via galaxy mergers rather
than in-situ star formation, then star forming galaxies with large HI
masses will be largely absent from the most massive dark matter
halos. HI selected satellite galaxies will also be rare if HI galaxies
merge rapidly after the merger of their host dark matter
halos. Similarly, if HI is consumed or removed from satellite galaxies
within dark matter halos, then satellite galaxies would be
under-represented in HI surveys relative to optical surveys as seems
to be the case based on our analysis of HIPASS. Although our HOD
results are suggestive of these scenarios, the precision with which
the HI HOD can be studied with HIPASS is limited. However the much
larger volumes that will become available with the advent of deeper HI
surveys such as those to be undertaken with the Australian SKA
Pathfinder (ASKAP, Johnston et al.~2008) will allow more detailed comparison of the halo
occupation of stars and HI. This will in turn facilitate formulation
of a more detailed understanding of the growth of stellar mass in
galaxies.

\section{Summary}
\label{summary}

In this paper we have analysed the clustering properties of HI
selected galaxies from the HIPASS survey using the formalism of the
halo occupation distribution. Use of the HOD model separates the
clustering amplitude into contributions from galaxy pairs that are in
the same halo (the 1-halo term) and pairs that reside in different
halos (the 2-halo term). The real-space clustering amplitude is
significant on scales below the virial radius associated with the halo
mass required to reproduce the clustering amplitude on large scales,
indicating that single halo pairs are contributing a 1-halo
term. However the resulting parameter constraints show that satellite galaxies make up only $\sim10\%$ of the HIPASS sample. HI satelite galaxies are therefore  less significant in number and in terms of their contribution to
clustering statistics than are satellites in optically selected galaxy
redshift surveys. Thus HOD modeling of HI galaxy clustering
quantifies the extent to which environment governs the HI content of
galaxies and confirms previous evidence that HI galaxies are
relatively rare in overdense environments (Waugh et al.~2002; Cortes et al.~2008). Through our
real-space modeling of HIPASS clustering we find a minimum halo mass
for HIPASS galaxies at the peak of the redshift distribution of
$M\sim10^{11}$M$_\odot$, and show that less than 10\% of baryons in
HIPASS galaxies are in the form of HI.

Our analysis reveals significant degeneracies in the HOD parameters
that give acceptable fits to the real-space HI correlation
function. However the extra line-of-sight dimension in the
redshift-space correlation function helps to break these degeneracies
because the fingers-of-god are sensitive to the typical halo mass in
which satellite galaxies reside. Our analysis of the redshift space
correlation function indicates that in order to get fingers-of-god in
a model which are as subtle as those in the HIPASS observations, the
HI rich satellites required to produce the measured 1-halo term must
be preferentially in group rather than cluster mass halos. In our
modeling the best representations of the fingers-of-god are obtained
by imposing an upper limit on the halo mass where HI satellites are
found of $\sim10^{13.7-14.3}$M$_\odot$. This finding is in accord with
direct observations of rich optical clusters, which show no
overdensity of HI galaxies relative to the field (Waugh et al.~2002; Cortes et al.~2008).
Quantitative constraints on HOD models from the HIPASS survey are
limited by the small survey volume, which makes the determination of
the correlation function systematically uncertain (Meyer et
al.~2007). Future deeper HI surveys with telescopes like the Australian SKA Pathfinder (ASKAP) will
survey a much larger volume (Johnston et al.~2008) and allow the distribution of HI with
environment to be studied in more detail via precise measurements of
clustering in HI galaxies.

The cosmic star-formation rate has declined by more than an order of
magnitude in the past 8 billion years (Lilly et al. 1996, Madau et al.
1996). The decline is observed across all wavelengths (Hopkins 2004
and references therein) and apparently defies observational
limitations such as sample selection and cosmic variance (Westra \&
Jones 2008).  Optical studies paint a somewhat passive picture of
galaxy formation, with the stellar mass density of galaxies gradually
increasing and an increasing fraction of stellar mass mass ending up
within red galaxies that have negligible star-formation (e.g., Brown
et al. 2008). However optical studies can only address part of the
picture.  Currently, the combination of direct HI observations at low
redshift (Zwaan et al. 2005b; Lah et al 2007) and damped Ly$\alpha$
absorbers in the spectra of high-redshift QSOs (Prochaska et al. 2005)
show that the neutral gas density has remained remarkably constant
over the age of the universe.  At these levels, and without
replenishment, HI gas would be exhausted in a few billion years
(Hopkins et al 2008). Models incorporating gas infall that balances
star formation and gas outflow are therefore necessary to reproduce
observed star formation densities (eg. Erb 2008). The evolutionary
and environmental relationships between the neutral gas which provides
the fuel for star formation and the stars that form are central to
understanding these and related issues. The study of the halo
occupation distribution of HI based on HIPASS galaxies presented in
this paper provides the first quantitative hints of this relationship.

\bigskip

\noindent
{\bf Acknowledgments.} The research was supported by the Australian
Research Council (JSBW).

\label{lastpage}
\end{document}